\documentclass[acmlarge]{acmart}
\DeclareUnicodeCharacter{0140}{\l}

\AtBeginDocument{%
  }

\setcopyright{acmlicensed}
\copyrightyear{2018}
\acmYear{2018}
\acmDOI{XXXXXXX.XXXXXXX}

\acmJournal{POMACS}
\acmVolume{37}
\acmNumber{4}
\acmArticle{111}
\acmMonth{8}

\usepackage{todonotes}

\begin{document}

\title{Navigating Decentralized Online Social Networks: An Overview of Technical and Societal Challenges in Architectural Choices}

\author{Ujun Jeong}
\affiliation{%
  \institution{Arizona State University}
  \city{Tempe}
  \country{USA}
}
\email{ujeong1@asu.edu}

\author{Lynnette Hui Xian Ng}
\affiliation{%
  \institution{Carnegie Mellon University}
  \city{Pittsburgh}
  \country{USA}
}
\email{lynnetteng@cmu.edu}

\author{Kathleen M. Carley}
\affiliation{%
  \institution{Carnegie Mellon University}
  \city{Pittsburgh}
  \country{USA}
}
\email{carley@andrew.cmu.edu}

\author{Huan Liu}
\affiliation{%
  \institution{Arizona State University}
  \city{Tempe}
  \country{USA}
}
\email{huanliu@asu.edu}

\renewcommand{\shortauthors}{Jeong et al.}

\begin{abstract}
Decentralized online social networks have evolved from experimental stages to operating at unprecedented scale, with broader adoption and more active use than ever before. Platforms like Mastodon, Bluesky, Hive, and Nostr have seen notable growth, particularly following the wave of user migration after Twitter’s acquisition in October 2022. As new platforms build upon earlier decentralization architectures and explore novel configurations, it becomes increasingly important to understand how these foundations shape both the direction and limitations of decentralization. Prior literature primarily focuses on specific architectures, resulting in fragmented views that overlook how different social networks encounter similar challenges and complement one another. This paper fills that gap by presenting a comprehensive view of the current decentralized online social network landscape. We examine four major architectures: federated, peer-to-peer, blockchain, and hybrid, tracing their evolution and evaluating how they support core social networking functions. By linking these architectural aspects to real-world cases, our work provides a foundation for understanding the societal implications of decentralized social platforms.

\end{abstract}



\received{20 February 2007}
\received[revised]{12 March 2009}
\received[accepted]{5 June 2009}

\maketitle

\section{\textbf{Introduction}}

Decentralized Online Social Networks (DOSNs) are platforms that distribute data and control across multiple independent entities, rather than relying on a single centralized authority~\cite{datta2010decentralized}. However, decentralization in the context of social networking involves more than just distributed infrastructure~\cite{ng2024smi}. A DOSN must support core social functions, including identity management, content distribution, and community governance, without centralized oversight. The degree of decentralization depends on the network’s architectural choice and its ability to foster trust, coordination, and user autonomy. To support these goals, early DOSNs adopted decentralization architectures such as peer-to-peer systems and federated networks operated by community-run servers~\cite{wan2024web3, masinde2020peer, sueur2012social}.

In recent years, DOSNs have moved beyond niche experiments and into platforms of significant interest. The emergence of new platforms, most notably \textit{Mastodon} and \textit{Bluesky}, has seen rapid growth in user registration and engagement. \textit{Mastodon}, for example, surged in popularity following upheaval at \textit{Twitter} in 2022 after the leadership and its major policy changes~\cite{jeong2024exploring, cava2023drivers, he2023flocking}, reaching approximately 15 million registered accounts by 2024~\cite{Guardian2024AltPlatforms}. Similarly, \textit{Bluesky} surpassed 25 million total users shortly after opening public registrations in late 2024~\cite{sahneh2024dawn}, having already attracted over a million active users during its beta phase~\cite{StatistaBluesky}. These platforms are publicly deployed and rapidly scaling, placing them under pressure to demonstrate whether decentralized platforms can sustain complex social ecosystems without introducing technical fragility or societal issues.

Despite this momentum, there remains a substantial gap in understanding how DOSNs operate, both among users and within academic literature. Many adopters arrive with assumptions shaped by centralized platforms like \textit{Twitter}, expecting similar approaches to moderation, identity, and visibility~\cite{ bono2024exploration, wang2024failed}. These assumptions often do not hold in decentralized environments. For instance, on \textit{Bluesky}, the list of who you block is public by default as a consequence of architectural design favoring transparency~\cite{kleppmann2024bluesky, schwittmann2013privacy}. Meanwhile, prior research has focused narrowly on early peer-to-peer or federated systems~\cite{koll2017good, chowdhury2015taxonomy}, offering limited insight into the architectural diversity and evolution seen in today's DOSNs. As a result, recent platforms remain largely unexamined, despite introducing novel design patterns and challenges. To name a few, \textit{Hive} employs blockchain consensus, while \textit{Nostr} uses cryptographic keys and relay-based communication~\cite{steem2018steem, wei2024exploring}. These emerging approaches introduce unfamiliar trade-offs such as data ownership, immutable public data, and decentralized moderation.

\vspace{0.25cm}

This work aims to reduce these gaps by helping social media researchers and adopters better understand the evolving DOSN landscape. Our main contributions are as follows:

\begin{itemize}
    \item \textbf{Providing a comprehensive architectural taxonomy and history:} We present the current taxonomy of DOSN architectures, including peer-to-peer, federated, blockchain, and hybrid models, along with their evolution and the technical challenges they have faced.

    \item \textbf{Revealing architectural limits to decentralization:} We examine how different architectures distribute core social data such as identity, content, interactions, and community to identify technical bottlenecks that impede seamless decentralized social networking.
    
    \item \textbf{Linking technical constraints to societal challenges:} We highlight how architectural choices contribute to enduring social challenges, including moderation gaps, spam, governance failures, and social fragmentation, even in the latest decentralized platforms.

\end{itemize}

The rest of this paper is organized as follows. Section~2 categorizes and traces the evolution of DOSN architectures, including federated, peer-to-peer, blockchain-based, and hybrid systems. Section~3 examines how these architectures manage the distribution and access of social data types. Section~4 explores the core technical trade-offs across architectures in terms of performance, security, and user control. Section~5 links these architectural foundations to broader societal challenges such as moderation, interoperability, and abuse prevention.

\section{\textbf{DOSN Architectures and History of Protocols}}
\begin{figure}[htbp]
    \centering
    \includegraphics[width=1.0\textwidth]{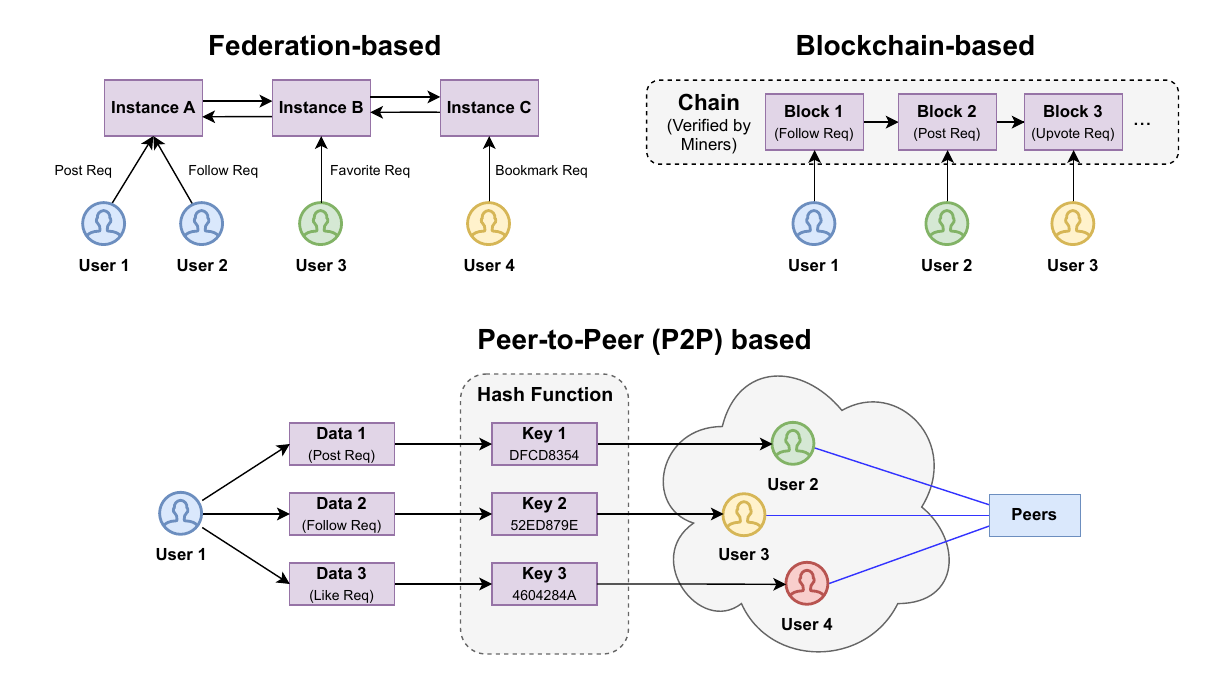}
\caption{High-level overview of the core mechanisms underlying DOSN architectures: federation, P2P, and blockchain models.}
    \label{fig:dosn_taxonomy_architecture}
\end{figure}

\begin{figure}[htbp]
    \centering
    \includegraphics[width=1.0\textwidth]{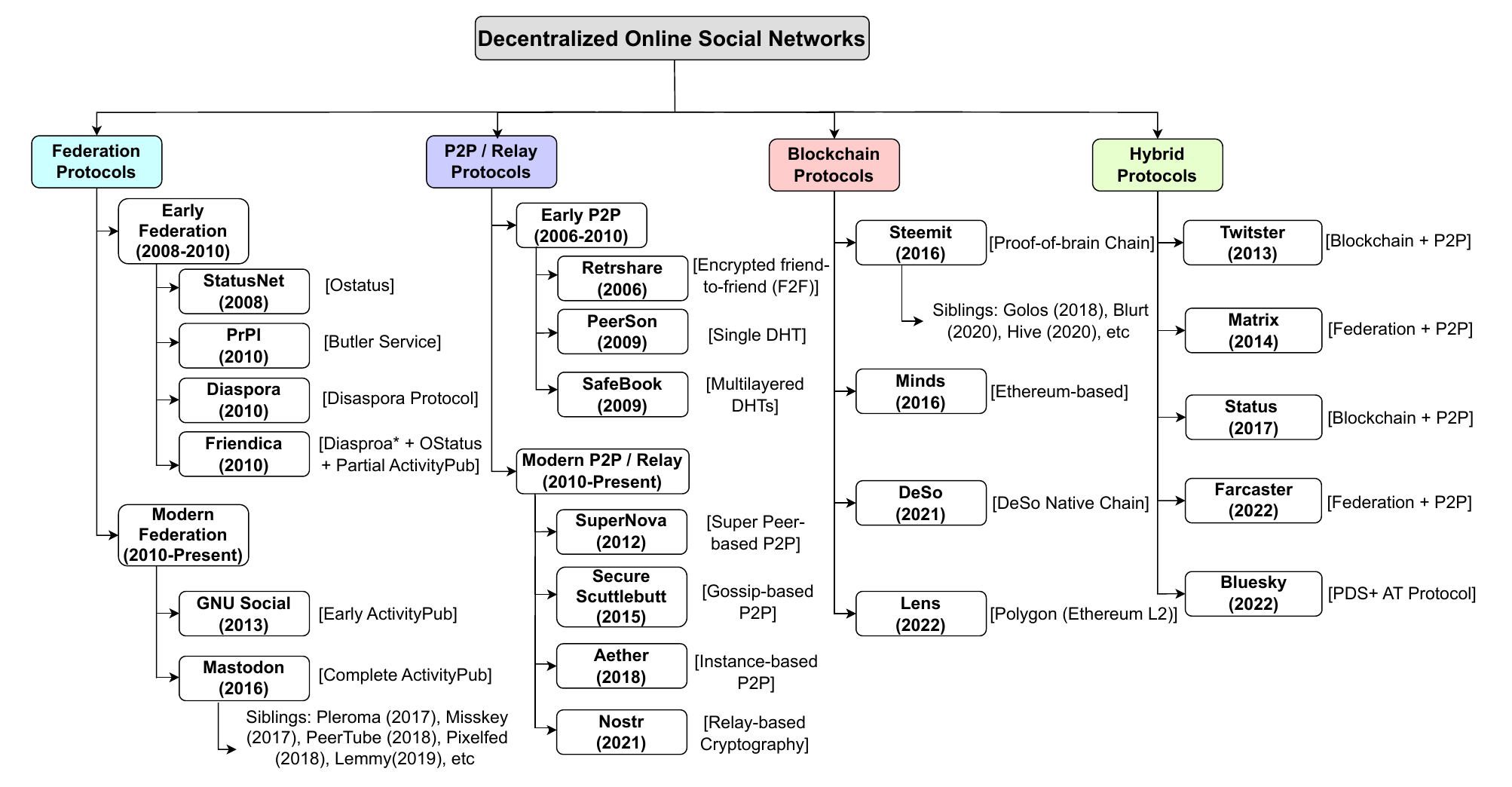}
    \caption{Evolution of projects across DOSN architectures. The projects are shown with the protocols on which they are built.}

    \label{fig:dosn_taxonomy_history}
\end{figure}

Architectures for DOSNs have undergone a broad expansion, driven by advancements in distributed systems, cryptographic identity, token economies, and the emergence of Web 3.0~\cite{marx2023decentralised}. While various DOSN architectures have been studied, a systematic and comprehensive introduction to recent developments remains lacking. Through an extensive review of the literature~\cite{riaz2022content, masinde2020peer, guidi2020blockchain, tapiador2018understanding, koll2017good, chowdhury2015taxonomy}, we identify four primary architectural categories: (1) Federation, (2) Peer-to-Peer, (3) Blockchain, and (4) Hybrid models. Figure~\ref{fig:dosn_taxonomy_architecture} presents a high-level technical overview of each architecture, and Figure~\ref{fig:dosn_taxonomy_history} illustrates their historical evolution alongside the protocol of projects.

\subsection*{\textbf{Federation Architecture}}
Federated social networks consist of multiple servers (often called instances or pods) that communicate through a standardized protocol. Instead of having one central hub, any individual or organization can run a server to host a subset of users. Each server operates independently with its own admins and policies, but servers exchange user content and interaction events to form an interconnected network of communities. This is analogous to email: users can communicate across servers, but no single server controls the whole system. Users thus have autonomy when they choose a home server that aligns with their preferences.



Federated approaches gained prominence in the early 2010s with pioneering platforms such as \textit{StatusNet}, one of the first decentralized microblogging systems. In this model, user data is stored on independently operated servers that act as intermediaries for data availability. Although these servers may not be fully trusted, data confidentiality is preserved through encryption and can only be accessed by users with the appropriate decryption keys. The \textit{PrPl} framework extended this idea by allowing users to run a personal ``butler'' service for data management, either on self-hosted machines or rented cloud infrastructure. \textit{Diaspora} reacted to a response to increasing public concern over privacy and centralized control. It introduced the concept of user-selectable pods, limiting federation to a known set of servers to reduce exposure to untrusted nodes~\cite{bielenberg2012growth}. Around the same time, \textit{GNU Social}, a successor to \textit{StatusNet}, continued to support OStatus protocol and laid the foundation for a more community-driven federated environment~\cite{ostatus2010}. \textit{Friendica} expanded on this by supporting multiple federation protocols --- including OStatus the Diaspora* protocol, and later ActivityPub --- thus acting as a bridge between otherwise disconnected networks. The need for standardization led to the adoption of \textit{ActivityPub} as a W3C recommendation~\footnote{\url{https://www.w3.org/TR/activitypub/}}. This enables a more interoperable federated ecosystem, exemplified by platforms such as \textit{Mastodon}, \textit{PeerTube}, \textit{Pixelfed}, and \textit{Misskey}, collectively referred to as the ``Fediverse''~\cite{la2021understanding}. This ecosystem allows users to maintain control over their data by hosting it on independently managed servers, while still supporting inter-server communication.

\subsection*{\textbf{P2P/Relay Architecture}}

A classic Peer-to-Peer (P2P) operates social networks without dedicated servers, relying instead on user devices to form a distributed overlay. Each user stores their own data locally and connects directly with others to share updates. These designs often employ Distributed Hash Tables (DHTs) to locate user profiles and content without centralized indexing~\cite{hassanzadeh2021dht}. DHTs provide a decentralized lookup mechanism by distributing key-value pairs across nodes, enabling efficient data retrieval by key without a central directory. Modern P2P/Relay or “semi-P2P” model incorporates relay nodes --- intermediaries that temporarily forward messages between peers without storing data persistently --- to support communication when peers are not simultaneously online~\cite{hei2006stochastic}.

Early P2P prototypes focused on decentralized models mainly for social networking. \textit{RetroShare} employed a friend-to-friend (F2F) encrypted overlay, connecting only mutually trusted peers—ensuring strong privacy but limiting scalability. \textit{PeerSoN} introduced a single-layer DHT for peer discovery and content exchange, supporting basic social features like messaging and profile sharing~\cite{buchegger2009peerson}. To enhance privacy, \textit{Safebook} proposed a multi-layered routing scheme where friends acted as proxies, masking identity and location through concentric trust circles~\cite{cutillo2009safebook}. However, these systems faced challenges with peer churn (i.e., the instability caused by frequent node turnover). In response to this, \textit{SuperNova} leveraged socially trusted peers as temporary relays, enabling content availability even when the original author was offline~\cite{sharma2012supernova}. Similarly, \textit{Secure Scuttlebutt (SSB)} used a gossip-based replication model optimized for offline-first scenarios, where each user’s feed is locally stored and incrementally shared~\cite{tarr2019secure}. \textit{Aether} takes a pseudonymous and ephemeral approach, distributing forum posts locally and applying community moderation through time-based decay and voting, instead of relying on persistent identity or central servers. Most recently, \textit{Nostr} has adopted a relay-centric design: clients sign and broadcast messages to a network of relay servers, and the servers forward the messages to subscribers without maintaining a global state.




\subsection*{\textbf{Blockchain Architecture}}
Blockchain has emerged as an alternative approach for decentralized social networks, enabling the recording of user interactions --- such as posts, likes, and follows --- as transactions on a public ledger. This ensures a globally consistent, tamper-resistant event log maintained by miners or validators instead of a central authority. By leveraging distributed ledger technology, such networks improve transparency, verifiability, and censorship resistance while addressing trust and governance challenges. However, due to high transaction costs and limited throughput, storing large media files fully on-chain has been often deemed impractical~\cite{guidi2020blockchain, pfeiffer1232020blockchain, poongodi2020blockchain}. For this reason, many blockchain-based DOSNs incorporate off-chain storage solutions, such as the InterPlanetary File System (IPFS) or Arweave, to handle large-sized data (e.g., videos, images, etc)~\cite{benet2014ipfs, williams2019arweave}.


Notable examples of blockchain that are publicly deployed include: \textit{Steemit}, its fork \textit{Hive}, and newer protocols like \textit{Lens}. \textit{Steemit} pioneered a “proof-of-brain” incentive model, rewarding users with tokens based on community voting~\cite{steem2018steem}, built atop a delegated “proof-of-stake” consensus. Several forks and clones --- including \textit{Hive}, \textit{Golos}, \textit{Blurt} --- replicate this model with variations in governance and tokenomics, often catering to niche communities or content themes. \textit{Lens} Protocol, by contrast, offers a composable social graph deployed as smart contracts on Polygon, a scalable chain compatible with Ethereum. In \textit{Lens}, core social interactions are tokenized as NFTs, while content is stored off-chain via decentralized storage like IPFS or Arweave. Other platforms like \textit{DeSo} and \textit{Minds} similarly aim to decentralize identity and content. \textit{DeSo} runs on a purpose-built blockchain optimized for social data, storing profiles, posts, and creator tokens directly on-chain.




\subsection*{\textbf{Hybrid Architecture}}
In our study, we define hybrid architecture as a model that blends at least two distinct DOSN models (e.g., P2P and blockchain) to balance scalability, persistence, data control, and administrative efforts. \textit{Twister} combines a Bitcoin-like blockchain for identity with a BitTorrent DHT for content sharing. \textit{Matrix} uses federated home servers for communication and encryption, with emerging support for peer-to-peer extensions. \textit{Status} integrates Ethereum identity with Waku, a gossip-based P2P messaging protocol, 
 to enhance private communication~\cite{thoren2022waku}. \textit{Farcaster} anchors identity on Ethereum while using federated hubs to relay content. \textit{Solid} introduces user-controlled data pods with global interoperability via linked data standards. Extensible Message Transport Protocol (XMTP)\footnote{\url{https://xmtp.org/}} further extends privacy by enabling End-to-End Encrypted (E2EE) messaging tied to blockchain keys. Most recently, \textit{Bluesky} has gained broad public traction with its AT Protocol, which introduces Personal Data Servers (PDS)\footnote{\url{https://atproto.com/guides/self-hosting}} --- independent hosts that form a federated ecosystem supporting data portability and interoperability.

\section{\textbf{Data Access in DOSNs}}
A modern social media platform handles a variety of data types, which can be broadly classified into four main categories: \textbf{Profile data}, \textbf{Content data}, \textbf{Interaction data}, and \textbf{Community data}. This section defines each category and explores how different decentralized online social network architectures manage their storage, access, and distribution. Although there are more granular data types, such as login data,  we focus on these four primary categories due to their central role in social interactions~\cite{richthammer2014taxonomy}. Definitions of data are presented in Table~\ref{tab:data-categories}.

\begin{table}[h!]\small
\centering
\caption{Data Categories in Online Social Networks}
\label{tab:data-categories}
\begin{tabular}{p{3cm}p{12cm}}
\hline
\textbf{Category} & \textbf{Description and Examples} \\
\hline
\textbf{Profile Data} & Information that defines a user's identity and presence within the network, including attributes such as ID, username, profile picture, bio, and demographic details. \\  
\textbf{Content Data} & User-generated content encompassing text posts, images, videos, and other media forms. \\  
\textbf{Interaction Data} & Records of social interactions and relationships among users, such as likes, shares, and mentions. \\ 
\textbf{Community Data} & Data pertaining to group behaviors, statistics, and community-level recommendations, including memberships, moderation rules, trending topics, and hashtags. \\  

\hline
\end{tabular}
\end{table}
\subsection{\textbf{Profile Data}}
Profile data encompasses the information that defines a user's identity within the social network. This includes unique identifiers (such as usernames or handles), display names, profile images, biographies, and additional metadata like account creation dates or verification statuses. In some cases, cryptographic identity keys are also incorporated to enhance security and authentication.

\subsubsection{\textbf{Federation Architecture}} Profile data is primarily managed by the user's home server --- the instance where they registered. For example, if a user \texttt{@Alice@ServerA} creates an account on \texttt{ServerA}, this server is responsible for storing and managing her profile information, including her display name, avatar, bio, and other metadata. Federated platforms such as \textit{Mastodon} and \textit{Misskey} use protocols like ActivityPub to facilitate profile discovery and data sharing across instances. When a user interacts with others on different servers, their profile data is retrieved and temporarily cached by those servers. This allows remote users to view Alice’s profile without requiring constant direct queries to her home instance. Profile discovery in federation often relies on the WebFinger\footnote{\url{https://webfinger.net/}} protocol. Federated architectures generally offer profile portability, meaning users can migrate between instances while retaining their social connections.


\subsubsection{\textbf{P2P/Relay Architecture}} Profile data resides on the user’s device and is propagated through the network. Identity is often tied to a cryptographic key pair rather than a human-readable username. For instance, \textit{SSB} identifies users by a public key associated with a nickname in clients, while \textit{Nostr} uses a public key as the user’s global identifier. Profile data is broadcast as signed messages or documents and distributed among peers. Depending on the platform, this distribution occurs through gossip protocols, DHT, or relays. With no central authority managing profiles, users must propagate their profile information to ensure it reaches their contacts. Even if a user goes offline, cached copies of the profile may persist among peers. The P2P model often lacks global user search, meaning discovering a profile requires its identifier or finding it via friend-of-friend connections.

\subsubsection{\textbf{Blockchain Architecture}} Profile data is implicitly tied to the user’s blockchain account or wallet. A user may be represented by an account address (e.g., an Ethereum address) or an NFT token serving as a profile. Basic profile attributes, such as username and bio, are typically stored in a distributed manner. For example, the \textit{Lens} profile NFT contains a reference to metadata that includes the user’s display name, profile picture, and about section. Similarly, platforms like \textit{Steemit} allow users to register human-readable account names on-chain and store profile attributes in a JSON object within the blockchain state. Since identity is recorded on-chain, it remains globally unique and portable across different front-end applications that interact with the blockchain. Decentralized identity schemes (DIDs)~\cite{reed2020decentralized} also provide a way to associate blockchain accounts with off-chain profile data, allowing for selective disclosure of information. As a result, blockchain-based profiles can often be used seamlessly across multiple applications (e.g., a \textit{Lens} profile can be accessed across various clients).

\subsubsection{\textbf{Hybrid Architecture}} The hybrid approach integrates decentralized identity systems with conventional data storage to achieve both portability and performance. DIDs or blockchain-based accounts provide unique, portable user identities, while traditional databases offer fast and efficient access to profile data. For example, \textit{Bluesky} assigns globally unique DIDs to users, allowing them to move between servers without changing their identifiers. Profile information is stored in a personal data repository through PDS, so users can maintain their identity continuously across different domains.

\subsection{\textbf{Content Data}}
Content data includes text-based posts (such as microblog entries or full-length articles), multimedia files (photos, videos, audio), and their accompanying metadata. Additionally, comments and replies are also categorized under content data due to their dialogue-centric nature.


\subsubsection{\textbf{Federation Architecture}} Content data is stored on the user’s home server and shared with other servers as needed. For example, when a \textit{Mastodon} user writes a post, that post is saved in the database of their home instance. If the user has followers on other instances, the home server will send the post to the followers' remote servers. Those servers may cache the content data to serve it to their local users (e.g., media attachments might be cached or proxied), but the authoritative copy resides on the origin server. Federation typically uses on-demand content replication: a post is copied to servers that have an interested audience --- such as followers or participants in a conversation --- but it isn’t distributed or stored across all servers in the network. Some federated platforms allow or require users to manually export/import their content for migration. For example, a user switching servers might take their archive of posts to the new server, but this process is not seamless, and old posts may not automatically appear on the new account. From an access perspective, retrieving content in a federated network usually involves querying the relevant servers. If you search for or request an older post not cached on your server, your server may have to fetch it from the origin server.



\subsubsection{\textbf{P2P/Relay Architecture}} Content data is distributed among the peers themselves. Typically, the author’s device keeps the original copy, and the network propagates it to other peers based on the sharing model. There are a few approaches:
(1) \textit{Eager replication:} content is pushed out widely. For example, in friend-to-friend P2P networks, when Alice posts something, the content might be sent directly to all her friends’ devices, which in turn forward it to their friends, etc., so the post spreads like gossip and many peers eventually store a copy. (2) \textit{On-demand retrieval:}  content is not proactively replicated everywhere, but any peer can request it from the network when needed. For instance, a peer might use a DHT to find which peer has a given content hash, or query the author’s peer when it is online. 
Large media files in purely P2P networks can be problematic, so many P2Ps either expect users to host them on their own or integrate with external P2P storage networks. For example, one could share an image by providing a magnet link or IPFS hash, which the interested peers then use to fetch the data from whoever has it. The deletion of content in P2P is tricky. Once content is out, it cannot be reliably cleaned up by all peers.  user can delete content on their device and stop sharing it, but other users who already have a copy might keep it indefinitely unless those peers are signaled to delete it at a protocol level.



\subsubsection{\textbf{Blockchain Architecture}} Handling of content data depends on what the blockchain stores. Because blockchains have limited capacity, many platforms store only pointers or summaries on-chain while keeping bulk content off-chain. For instance, \textit{Hive} stores text content (posts/comments) on-chain as part of custom operations in blocks. This means a textual post becomes part of the permanent ledger; every full node stores the post and the post cannot be later erased. Media (images/videos) are not stored on the chain; instead, users typically upload those to IPFS or a centralized image host and include an IPFS hash or URL in the post. In Ethereum-based social protocols like \textit{Lens}, typically the content itself is not stored directly in Ethereum; instead, \textit{Lens} might store a content hash and reference on-chain, with the actual content hosted on IPFS. As long as someone is seeding the content in IPFS or decentralized persistent storage, the content can be fetched by anyone with the hash. The entire network sees a consistent view of published posts, which simplifies global discovery (e.g., a trending page can be built by scanning the blockchain for popular posts). Communities or the platform maintain servers to pin popular content or rely on users to cache what they care about. In other words, the blockchain approach regards content as a public commons: stored by many, accessible to all, and essentially permanent unless the whole network agrees to purge it, which is extremely rare.


\subsubsection{\textbf{Hybrid Architecture}} A common hybrid approach combines a decentralized index for content metadata with centralized or federated storage for efficiency. For example, \textit{Farcaster} records an immutable log of user actions on Ethereum to maintain a globally consistent sequence of posts, while the message content is stored on \textit{Farcaster}’s servers. If those servers go offline, the content may become temporarily inaccessible, but the on-chain record allows recovery from any peer with the data. \textit{Mastodon} also follows a hybrid model for media storage, with servers using a mix of local and remote storage, often through cloud-based infrastructure, and coordination handled through federation. A potential hybrid design might rely on a few stable servers to host large content like video while distributing lighter data such as text and metadata over P2P networks. This is the approach used by federated video platforms like \textit{PeerTube}, where users can delete content from primary storage, although backups may still exist across the distributed network.

\subsection{\textbf{Interaction Data}}
Interaction data encompasses records about how users react to the content or other users. This includes actions like likes/upvotes, shares/reposts, as well as relationship data such as follow/friend links. Interaction data ties together content and the social graph: it tells us who engaged with what or whom.


\subsubsection{\textbf{Federation Architecture}} Interaction data is managed by propagating events between the appropriate servers. For instance, if Alice (on \texttt{ServerA}) follows Bob (on \texttt{ServerB}), \texttt{ServerA} sends the following request to \texttt{ServerB}. If Bob accepts the request --- or if the platform allows open following by default, as in \textit{Mastodon} --- \texttt{ServerB} records Alice as a follower and subsequently delivers Bob’s updates to \texttt{ServerA}. The following relationship is stored on both sides: \texttt{ServerA} knows that Alice follows Bob, and \texttt{ServerB} knows that Alice follows Bob. This duplication ensures that each server can operate independently for the view of the world of its own users. Similarly, if Alice (on \texttt{ServerA}) likes Bob’s post (on \texttt{ServerB}), \texttt{ServerA} notifies \texttt{ServerB} of the like. \texttt{ServerB} will record the likes (e.g., increasing the like count and possibly storing Alice’s ID in the list of who liked the post), and \texttt{ServerA} could also keep a record. In federated networks, no single server has the entire interaction. Each server has a partial view: each server's local users and the connections those users have (local-to-local or local-to-remote). For example, there is no official global query to get all followers of a given user except to ask the home server of that user, because that server only knows the followers it has seen.

\subsubsection{\textbf{P2P/Relay Architecture}} Interaction data is part of gossip among peers. A follow can be implemented by exchanging keys or by each user adding the other to a locally stored friends list which is then shared. Because there is no server to notify the other parties, the act of following could be an event that gets published in the P2P network. For example, in \textit{SSB}, following someone means that you announce a ``follow” message --- digitally signed by you and referencing the other person's public key. That message propagates through the network. Any user that has both you and the target in their social horizon can learn about the connection. The following relationship is typically recorded in each user's profile or activity log. Some experimental P2P networks may choose to encrypt this information or share it selectively --- for example, by notifying each friend individually that you follow them, without publishing a public list of all your followers. Similarly, likes can be implemented as signed messages referencing the content’s ID, which are distributed among peers. Counts like ``total likes on a post” are just aggregated from what each user has seen. A user's device might compile how many like messages it has encountered for a given post, but another user’s device might have seen a different set. P2P systems may not even try to maintain global counts, showing instead only the interactions from people you directly know, to avoid needing a full-network view (e.g., ``5 of your friends liked this” rather than ``5000 people liked this”).

 
\subsubsection{\textbf{Blockchain Architecture}} In blockchain-based social networks, interaction data can be recorded directly on-chain. For example, if Alice likes Bob’s post, this may be represented as a blockchain transaction --- such as a custom ``upvote" operation. On \textit{Steemit}, votes were on-chain events that influenced a post’s score and reward; follow and unfollow actions were also recorded as specialized on-chain operations, making the social graph publicly accessible on the blockchain. Similarly, in \textit{Lens}, following a user triggers the issuance of a ``follow NFT," a token representing that relationship. As a result, the entire follow graph resides on-chain and is publicly viewable. Upvotes are also typically recorded on-chain. However, storing every interaction on-chain can lead to ledger bloat. To mitigate this, some platforms like \textit{Steemit} offer lightweight interactions (e.g., voting) free of micro-transaction fees, or opt to move certain interactions off-chain. For instance, the Mirror blogging platform on Ethereum does not store ``likes" on-chain; instead, users can express appreciation by collecting their NFT.

\subsubsection{\textbf{Hybrid Architecture}} In the context of blockchain, interaction data can be divided between on-chain and off-chain components. A hybrid approach might, for example, store the social graph (such as follower relationships) on-chain or in a globally accessible service to enable discoverability while keeping ephemeral interactions --- like likes or views --- off-chain to reduce clutter and enhance privacy. In one model, follows could be publicly visible on-chain, whereas likes are stored only on the content’s home server and not shared globally. Alternatively, the model could be reversed: follow data might be distributed P2P, with each user announcing who they follow, while a centralized service aggregates and computes metrics like trending hashtags or popular posts. In the context of federation, \textit{Bluesky} exemplifies a hybrid model where multiple algorithmic feeds can coexist, allowing users to choose feeds based on criteria such as engagement levels. In this approach, likes and reposts may be aggregated by the algorithm rather than the network itself. This represents a hybrid design where raw interaction data remains distributed, while a semi-centralized service compiles it into personalized feeds. Another hybrid concept involves selective disclosure, where a user might like a post privately so that only the author is notified, or publicly so that the like count is visible to all.



\subsection{\textbf{Community Data}}
Community data includes aggregated or high-level information about the social network, such as membership details, group descriptions, trending topics and hashtags, usage analytics, shared moderation resources (e.g., lists of blocked servers or users), and various other statistics reflecting collective user behavior.

\subsubsection{\textbf{Federation Architecture}}

Community structures in federated systems are often present but tend to be ad hoc or locally defined. For instance, \textit{Mastodon} does not include a built-in notion of “communities,” yet users have organically developed practices such as community accounts or the use of hashtags to foster topic-based social grouping. Other federated platforms, such as \textit{Friendica} and \textit{Diaspora}, offer features like aspects or forums that allow users to join shared spaces. These may be implemented under the hood as special accounts or as posts distributed among all group members. Membership in such groups is essentially inferred from interaction data—such as a list of participating users—but managing this requires coordination. To establish a group that spans multiple servers, one server may act as the “home” of the group while others subscribe to its content. Trending topics are typically computed at the instance level. For example, each \textit{Mastodon} instance can independently determine which hashtags or posts are trending among its own users. Similarly, metrics such as active user counts or daily post volume are tracked locally and are sometimes aggregated manually by third-party tools. Moderation decisions, such as blocking server domains, are handled autonomously by each server’s admins. However, many servers publish their block lists publicly, and others may adopt or adapt them. This practice leads to a loosely coordinated, community-driven dataset of flagged domains.



\subsubsection{\textbf{P2P/Relay Architecture}} In a pure P2P architecture, collecting or maintaining community-level data is inherently difficult. Without centralized servers to aggregate information, identifying global trends or computing network-wide statistics depends on either voluntary reporting by individual nodes or extensive crawling across the network. As a result, P2P social networks tend to form around small, localized social circles. Users are aware of what their friends are discussing, but there is no inherent mechanism for discovering global conversations or building a large-scale community feed. P2P networks like \textit{SuperNova} employ super peers as volunteer users who run enhanced nodes or lightweight servers that take on additional responsibilities. These super peers can aggregate community-level data, help disseminate popular content, or serve as discovery hubs, enabling features like trending topics or community statistics without introducing full centralization


\subsubsection{\textbf{Blockchain Architecture}} Blockchain inherently has a global view of data, so community data can often be handled easily. If all posts and interactions are on one ledger, one can compute trends and statistics by analyzing the chain. Some blockchain socials explicitly include community features: for example, \textit{Hive} has a ``Communities” feature where a community is like a subreddit --- this feature is implemented as a special account that users post to. Membership might be implied by who follows the community account or who has posted in it. Because all of that is on-chain, one can retrieve a list of members or content in the community by querying the blockchain state. Trending topics can be computed by scanning recent blocks for most-used tags or upvoted content. Not all community-relevant information needs to reside on-chain. Functions like content recommendations or personalized trends are typically handled off-chain by front-end algorithms. For instance, a moderator in a blockchain-based forum might maintain an off-chain list of banned users, ensuring their posts are hidden from the community feed. However, the blockchain itself may not support a direct concept of banning, since it cannot prevent anyone from writing to the chain. That said, blockchain systems can still support community governance through smart contracts. A community might issue its own token and use token-weighted voting to elect moderators or establish rules, with decisions enforced through on-chain proposals.



\subsubsection{\textbf{Hybrid Architecture}} In a hybrid P2P model, community data can still be managed in user-friendly ways by introducing dedicated servers or overlay networks that act as community coordinators. These coordinators track membership, aggregate relevant content, and facilitate communication across multiple P2P nodes. For example, an open-source server might maintain a list of members within an interest group, collect tagged posts, and generate a group timeline or content summary. If such a server becomes unavailable, the data may still persist with individual members --- particularly when direct P2P replication is supported. Platforms such as \textit{SSB} and \textit{Matrix} support shared channels that serve as community hubs, structured through DHTs or coordinated through designated peers. In blockchain-based contexts, a distributed ledger or consensus among a subset of community leaders can be used to maintain a shared state --- such as membership lists, rules, or moderation decisions. This approach can be more lightweight and flexible than relying on a global blockchain. A shared state may take the form of a mini-ledger or a Conflict-free Replicated Data Type (CRDT)~\cite{shapiro2011conflict}, synchronized only among interested participants, thereby preserving decentralization while enabling coherent group functionality.




\section{\textbf{Technical Trade-offs across DOSN Architectures}}
Each DOSN architecture entails inherent trade-offs in its capabilities, primarily driven by how data access is accessed and managed~\cite{schneider2019decentralization}. We studied four key technical dimensions of the architecture: (1) Scalability, (2) Persistence, (3) Data Protection, and (4) Administration. These dimensions are selected based on extensive review and harmonization from literature~\cite{riaz2022content, masinde2020peer, guidi2020blockchain, tapiador2018understanding, koll2017good, chowdhury2015taxonomy}.
Table~\ref{tab:tradeoffs} provides an overview of these trade-offs.
\renewcommand{\arraystretch}{1.3} 
\begin{table*}[h!]\small
\centering
\caption{Comparison of Technical Trade-offs of DOSN Architectures Across Key Dimensions}
\label{tab:tradeoffs}
\renewcommand{\arraystretch}{1.4}
\begin{tabular}{p{2.8cm}p{2.8cm}p{2.8cm}p{2.8cm}p{2.8cm}}
\hline
\textbf{Dimension} & \textbf{Federation} & \textbf{P2P/Relay} & \textbf{Blockchain} & \textbf{Hybrid} \\
\hline

\textit{Scalability} 
& Grows via independent servers; may face bottlenecks on large or under-resourced instances.  
& Serverless design enables autonomy; lookup delays and churn impact responsiveness.  
& Consensus limits throughput; requires off-chain support for real-time use.  
& Balances performance and decentralization with selective use of central services. \\

\textit{Persistence} 
& Content depends on the host server; loss risk if the instance shuts down without backups.  
& Data availability varies with peer uptime; ephemeral or unpopular content may vanish.  
& On-chain data is durable and tamper-resistant; large files are often stored off-chain.  
& Mixes server and peer storage; enables migration and fallback without full reliance. \\

\textit{Data Protection} 
& Instance admins can access user data; E2EE and metadata protection are limited.  
& Strong encryption and no central trust; IP and relay metadata may leak.  
& Public by design; secure but transparent; identity tied to private keys.  
& Sensitive data is often encrypted; users manage keys and choose trusted hosts. \\

\textit{Administration} 
& Instance-level rules; users can migrate, but moderation is uneven.  
& No global governance; moderation occurs locally or via clients/relays.  
& Immutable data; moderation handled by UI filtering or blacklists.  
& Supports user mobility with pluggable moderation and provider choice. \\
\hline
\end{tabular}
\end{table*}

\subsection{\textbf{Federation Architecture}}

\subsubsection{\textbf{Scalability}} Federated networks scale by adding independently operated servers, allowing the platform to grow horizontally. This architecture has demonstrated the capacity to support millions of users, but large instances may experience load challenges and require significant resources to maintain. A core trade-off lies in economic sustainability: while server-based systems can enable high availability of user data with reasonable performance, they often struggle to do so in a technically and economically feasible way without relying on externally centralized services. Many federated servers are run by volunteers, which leads to uneven reliability --- some instances may become overloaded, go offline, or shut down entirely due to limited resources.

\subsubsection{\textbf{Persistence}}
Consistency on data in a federated system is only as strong as the server that hosts a user’s content. If an admin shuts down their instance, the data on that server --- posts, media, and profiles --- may be lost, unless users have backups or the content has been cached by others. Unlike centralized platforms, there is no universal data replication across the network. Although followers in other servers might retain some content, there is no guarantee of long-term availability, making content persistence fragile and inconsistent.

\subsubsection{\textbf{Data Protection}} Sensitive data (e.g., profiles, user-generated content, and social graphs) is distributed, but not guaranteed to be protected. Federation avoids central control, which can enhance autonomy, but users must still trust their server admins, who have access to sensitive data. End-to-end encryption is limited, and metadata --- such as IP addresses or activity logs --- can still leak. Privacy protections vary widely from one instance to another, and users often lack visibility into how their data is handled behind the scenes.
\subsubsection{\textbf{Administration}} 

Administration is carried out at the instance level, allowing each server to define and enforce its own rules. This decentralized model supports community-driven governance and ideological diversity, but often leads to inconsistent moderation practices across the network. To address this, many instance admins participate in shared moderation frameworks, using community-maintained tools such as the “Oliphant Social Blocklist”~\cite{OliphantBlocklist} and “The Bad Space Garden”~\cite{GardenFence} to coordinate responses to controversial behavior or abusive servers. While no central authority can censor the entire system, collective action through sharing the list of defederation isolates controversial communities. For example, when some \textit{Mastodon} instances associated with the controversial forum \textit{Kiwi Farms} appeared, many admins acted in concert to defederate from it~\cite{Caelin2022}. 
 %
\subsection{\textbf{P2P/Relay Architecture}}

\subsubsection{\textbf{Scalability}} Pure P2P architectures such as early research prototypes like \textit{PeerSoN}~\cite{buchegger2009peerson} and recent protocols like \textit{Nostr} or \textit{SSB} do not require dedicated servers, instead having users’ devices form an overlay network. Each user stores and shares data, cooperating to forward messages and content. This approach maximizes user autonomy and privacy --- there is no need to trust a server or admin with one’s data or identity keys. It also excels in censorship resistance, as content can propagate as long as some peers remain online. For instance, the \textit{Nostr} protocol uses relays --- lightweight message servers that anyone can run. Users can publish to multiple relays to ensure their posts persist and are hard to suppress. If one relay blocks or drops content, interested users can retrieve it from another, and no central authority can globally ban the user. However, the trade-offs come in scalability and performance. P2P networks face challenges in delivering the responsiveness expected from centralized services. Data retrieval can be slow, as content may need to be fetched from intermittently online peers or from distant nodes via several hops in a DHT. Studies have noted lookup latencies on the order of seconds for P2P social queries, far higher than client/server architectures.

\subsubsection{\textbf{Persistence}} Ensuring high availability of data is difficult in P2P systems when many users go offline frequently. For example, on mobile devices, if a user’s friends are all offline, their data might not be immediately accessible. Recent work measuring \textit{Nostr} relays found that while decentralization is strong, relay availability can be an issue, partly due to the cost burdens on those running them~\cite{steem2018steem, wei2024exploring}. To ensure reliability, some \textit{Nostr} relays have begun charging small fees, primarily as a barrier to entry, preventing fake accounts from being involved.

\subsubsection{\textbf{Data Protection}} The appeal of P2P lies in its zero-trust design, where the final data sovereignty lies with the user. Features like end-to-end encryption, user-provided storage, and censorship resistance are inherent~\cite{masinde2020peer}. Since there is no centralized server or admin, users do not have to place trust in any third party to safeguard their identity or data. This model offers strong privacy guarantees and resilience to surveillance or control.

\subsubsection{\textbf{Administration}} P2P networks are highly resilient, with no single point of failure, and they empower users with autonomy. However, they depend on altruistic resource sharing and often sacrifice the speedy, always-on experience users expect. Administration is decentralized and difficult to coordinate. There is no universal enforcement mechanism, and moderation must occur at the client or relay level. Many P2P social systems remain experimental for these reasons, or they adopt partial P2P models to improve usability.

\subsection{\textbf{Blockchain Architecture}}

\subsubsection{\textbf{Scalability}} Blockchain-based social networks use a ledger to store social data or at least social proofs such as transactions representing posts, likes, and other actions. Examples include \textit{Steem} blockchain, and \textit{Minds}, which integrates Ethereum for token rewards. Scalability is a known challenge in this architecture. Public blockchains have limited throughput and high latency due to the need for global consensus. \textit{Steemit} mitigated this limitation using delegated proof-of-stake with a limited number of validators, achieving better throughput than Bitcoin or Ethereum~\cite{li2019incentivized}. Nonetheless, it still falls short of supporting large-scale, high-frequency platforms like Twitter without relying on centralized infrastructure or layer-2 enhancements.

\subsubsection{\textbf{Persistence}} One of the core strengths of blockchain architecture is data persistence. Once recorded on-chain, data is replicated across all nodes and becomes effectively immutable. This ensures strong guarantees of tamper-resistance and long-term availability. Users do not have to worry about admins silently deleting or altering their posts, as deletion is intentionally difficult. For example, \textit{Steemit} made user posts and upvote records immutable, meaning they could not be retroactively censored or doctored. As long as the blockchain continues to be maintained by a decentralized community of miners or validators, the content remains accessible and durable.

\subsubsection{\textbf{Data Protection}} In theory, blockchain networks offer high user autonomy. Users manage their identity through private keys and can often access the same data across different compatible front-ends (e.g. \textit{Steemit}, \textit{Hive}, etc). There is no centralized authority capable of removing a user's data at the protocol level. Instead, community-based mechanisms --- such as consensus-driven blacklisting --- are reserved for extreme cases like fraud or attacks. However, this autonomy comes with privacy and security trade-offs. If a private key is lost or compromised, there is no means of account recovery. Moreover, on-chain data is publicly visible and often linkable to user activity over time, leading to potential metadata exposure.

\subsubsection{\textbf{Administration}} Administration in blockchain-based systems operates at the application layer. While the underlying ledger preserves all data immutably, front-end interfaces can choose which content to display. For example, \textit{Steemit}’s website may hide certain content --- such as offensive material --- even though it remains permanently available on the blockchain. This results in a bifurcated model: strong protocol-level censorship resistance combined with necessary front-end moderation. Thus, while no single actor can delete content from the network, responsible moderation practices still emerge through user interfaces and community norms.


\subsection{\textbf{Hybrid Architecture}}

\subsubsection{\textbf{Scalability}} Hybrid architectures aim to combine the strengths of both decentralized and centralized systems by integrating P2P elements with central servers or cloud services. Many academic systems fall into this category, using servers for tasks such as fast indexing, search, bootstrapping, and heavy data storage, while decentralizing the control of social data. For example, \textit{PrPl} and \textit{Persona} allowed users to store encrypted data on personal or cloud servers but used a central index service to help users discover each other’s data. Similarly, \textit{Confidant} enabled P2P-style storage sharing among friends while relying on a central server for naming and discovery. These designs can support large-scale systems by offloading performance-intensive tasks to servers while preserving decentralization in key areas, although reliance on central components can still introduce scalability bottlenecks or single points of failure.

\subsubsection{\textbf{Persistence}} Hybrid systems enhance data availability by incorporating reliable server infrastructure without fully centralizing data ownership. Servers in these architectures often serve as intermediaries — storing encrypted data or routing messages — but are designed to be easily replaceable or minimal in trust. For instance, in systems like \textit{Persona}, users control the encryption keys while relying on cloud or personal servers for storage. In Confidant, friends stored each other’s data redundantly. More recently, \textit{Bluesky}’s AT Protocol enables users to select a PDS to host their data. If a PDS becomes unavailable or untrustworthy, users can migrate to another provider without losing access to their identity or content~\cite{kleppmann2024bluesky}. This flexibility supports data persistency while avoiding lock-in to any single provider.

\subsubsection{\textbf{Data Protection}} Hybrid architectures often aim to minimize trust in centralized components by keeping sensitive data encrypted and under user control. As noted in prior work~\cite{koll2017good}, hybrid systems can offer high user data availability through the use of servers, while still keeping communication and data private among users. For example, a server may route messages or store ciphertext but cannot access the underlying content. Users typically retain their own encryption keys, ensuring that only intended recipients can access their data. This model combines the reliability of server infrastructure with the confidentiality of P2P communication. In \textit{Bluesky}, decentralized identities and portable domains further reinforce user autonomy over data.

\subsubsection{\textbf{Administration}} Administration in hybrid systems varies depending on the design but often involves some centralized or community-maintained services. For example, while \textit{Bluesky} users control their identities and can switch between hosting providers, the ecosystem still includes federated network indexers responsible for content moderation~\cite{kleppmann2024bluesky}. These indexers are not protocol-mandated but form part of the shared infrastructure expected by users. This introduces a trade-off: while users are not locked into a single provider and can move their identities, the broader system still relies on the availability of its centralized components.

\section{\textbf{Societal Challenges of DOSNs}}
In this section, we examine how decentralization can lead to various societal challenges, breaking them down into six key areas: (1) Platform Safety, (2) Data Sovereignty, (3) Interoperability, (4) Operational Sustainability, (5) Account Credibility, and (6) Behavioral Dynamics. By drawing on real-world cases, we show how many of these social issues are related to the underlying architecture of decentralized social networking systems.

\subsection{\textbf{Platform Safety}}
Platform safety refers to the enforcement systems that prevent abuse, ensure timely rule application, and maintain legal compliance across decentralized social networks.


\subsubsection{\textbf{Moderation Responsiveness}}

 In federated systems, moderation is handled independently by server admins, leading to inconsistent enforcement—some apply strict rules and block entire instances, while others take a more permissive stance. Tools like \textit{Mastodon}’s “domain block” help admins reject content from problematic servers and form trust-based alliances, creating semi-autonomous, self-governing clusters. However, this also enables “moderation shopping,” where users evade enforcement by migrating to more lenient servers, and places a heavy burden on volunteer admins managing inappropriate content with limited resources. P2P networks lack any central authority and instead rely on trust-based content filtering, where users only see posts from those they trust. While this can reduce exposure to abuse, it shifts responsibility to individuals and complicates collective action. Reputation systems may help flag malicious actors, but they are hard to implement without centralized infrastructure and are vulnerable to Sybil attacks, where one user controls many fake identities~\cite{zhang2014sybil, viswanath2010analysis}. Governance in P2P networks is slow, dependent on social consensus and software updates. Blockchain-based platforms attempt to hard-code moderation using incentives or voting mechanisms; for example, \textit{Steemit} allows users to downvote content, reducing its visibility and rewards. While this crowdsourced approach reflects community sentiment, it is easily manipulated by users with large stakes~\cite{de2019fragility} and often prioritizes popularity over safety or well-being. Emerging hybrid models, such as \textit{Bluesky}, offer user-customizable moderation through third-party services and algorithms, recognizing that no single policy suits all communities. Although ATproto aims for user-driven advancement, progress has been slow because it was originally created by a corporation, not an open standards organization like the W3C, which typically fosters broader open-source support~\cite{kleppmann2024bluesky}.


\subsubsection{\textbf{Illegal Content}}

Handling illegal content --- especially cases like Child Sexual Abuse Material (CSAM) --- is a major challenge in decentralized systems. In federated architectures like the Fediverse (e.g., \textit{Lemmy}, \textit{Pixelfed}, etc), software is independently hosted and modified by admins to meet local needs, making it difficult to impose standard moderation tools akin to major platforms. This results in varied content policies; some servers permit adult material with minimal safeguards such as tags or warnings. A notable case is \texttt{pawoo.net}, a Japanese server criticized for hosting anime content depicting minors in sexualized contexts. Though legal locally, it conflicted with norms elsewhere, leading to widespread defederation and highlighting the difficulty of enforcing shared standards across loosely connected communities. The Stanford Digital Repository warned that 112 \textit{Mastodon} instances, including \texttt{mastodon.xyz}, were found sharing CSAM during a two-day analysis, yet moderation tools like PhotoDNA remain optional and inconsistently used~\cite{thiel2023child}. P2P systems exacerbate moderation challenges by removing servers and admins, allowing content to circulate directly between users without oversight, logging, or control—making it nearly impossible to detect or remove illegal material. Blockchain platforms present an even greater concern: once harmful content is embedded, it becomes permanently accessible due to the immutability principle, with no option for deletion~\cite{gibbs2018child}. Unlike federated or P2P systems where some intervention is possible, blockchains inherently prevent removal, raising ethical, legal, and technical issues for user-generated content.



\subsubsection{\textbf{Scamming}}
Scammers often exploit the lack of centralized verification on decentralized platforms. On \textit{Mastodon}, for example, users have reported scammers posing as admins or support staff, asking for passwords --- classic phishing tactics. New users may not realize that official announcements only come from the admin’s account on the same instance, making fake DMs seem credible. Crypto-related scams are also vital. Platforms like \textit{DeSo}, or even individual users discussing crypto, can attract fraudsters promoting fake investments or phishing links~\cite{sec2024nader}. Without a central anti-fraud team, users must largely fend for themselves. Some communities share warnings, but these are not always sufficient. The close-knit feel of smaller instances can also foster a false sense of security (“we’re both on this niche server, so I can probably trust them”). This makes social engineering especially effective, such as in dating scams where scammers build trust before luring users off-platform~\cite{flanagan2024nigerian}. Most critically, when scams occur, it is often unclear who is responsible or whom users can contact for support.
\subsection{\textbf{Data Sovereignty}}
Data sovereignty refers to the ability of users and admins to retain control over how data is stored, accessed, and secured within a decentralized environment.

\subsubsection{\textbf{Privacy}}

Federated networks often expose more user data than participants might expect. For the sake of interoperability, user profiles and posts are typically accessible via public APIs or open repositories. Users may assume that a smaller platform offers more privacy, but without siloed or hidden data structures, content is easily crawlable. For example, \textit{Mastodon} posts on public instances are viewable across the network and in web browsers, making them accessible to researchers, companies, or third-party aggregators. Although some posts can be marked “unlisted,” they may still appear in public databases. Direct messages on \textit{Mastodon} are not end-to-end encrypted and are stored in plain text on servers. While not publicly visible, they are accessible to admins. The decentralized model also means inconsistent security practices across instances: instead of protecting one large platform, the burden is distributed across thousands of smaller servers, some of which are hobbyist-run and poorly secured. This has led to cases of misconfigured instances leaking user data. Privacy in P2P networks raises similar concerns. Data are shared directly among nodes without a central authority, often exposing metadata such as social connections, timestamps, or interaction patterns to peers in the network. While some P2P systems attempt to mask origin or destination pseudonymous identities, robust encryption and anonymity are not always implemented~\cite{pouwelse2008tribler}. Blockchain-based architectures pose even greater privacy challenges. Since blockchains are typically public and immutable, all transactions and interactions are permanently visible. Many blockchain-based identity systems, such as those built on Polygon ID, integrate zero-knowledge proofs or selective disclosure mechanisms to enable verification without revealing the underlying data~\cite{zyskind2015decentralizing}.


\subsubsection{\textbf{Data Ownership}}
Data ownership is improved in the sense that users or communities host their data, but this also means if they fail to do so reliably, data can disappear. Also, if a server admin decides to shut down and not give data out, users could lose access to their own posts (which might be seen as a privacy violation or at least a loss of agency). In P2P cases, if a user’s device is the sole holder of some data and if the device is lost, the data is also gone. DOSNs often allow users and communities the right to decide what to do with their data. For example, a community could choose to purge its data after some time (e.g., some \textit{Diaspora} pods allowed users to auto-delete old posts). In a centralized network, users are at the platform’s mercy for such features. In a decentralized one, users can at least delete local copies and request others honor deletion. Federated networks allow users to export their data and migrate, which is a form of sovereignty --- you are not locked into a platform. Another emergent issue is web scraping and AI training. Because many federated networks' content is accessible via open APIs, there’s nothing to stop AI companies from scraping it to train large language models. Unlike Twitter, there is no central Terms of Service (ToS) to prohibit scraping. Users might feel a sense of violation if their content is used in AI without consent. This is a broad internet issue, but DOSN users tend to be more sensitive to it given their reason for being there.



\subsection{\textbf{Interoperability}}
Interoperability describes the ability of independently developed servers and protocols within a decentralized network to communicate, exchange content, and maintain consistent social features.


\subsubsection{\textbf{Social Fragmentation}}

The fediverse addresses fragmentation by encouraging standard protocols so that many small communities still form a larger federated network. The success of \textit{Mastodon} and others platforms in late 2022 showed that users could jump ship from a central platform and still remain interconnected on a decentralized network. However, there are still fragmentation points: \textit{Mastodon} vs. other federated protocols (like some networks use \textit{Disaspora} protocol or \textit{Nostr}, both of which do not connect to ActivityPub). Bridging these is non-trivial; some gateways exist (e.g., relays that cross-post between Nostr and \textit{Mastodon}, or plans to have \textit{Mastodon} read \textit{Bluesky} posts via AT Protocol), but it is evolving. When a centralized service like \textit{Twitter} faces controversy, many users diaspora can dilute network effects --- some friends go to \textit{Mastodon}, others to \textit{Bluesky}, etc. If those networks do not connect, social graphs fracture. In late 2022, we saw users juggling multiple accounts or “dual-wielding” \textit{Twitter} and \textit{Mastodon}~\cite{jeong2024exploring, jeong2024user}. Such dynamics are an inherent risk of decentralization, but one that can be managed with open standards and bridges. It might also be seen as a feature: it allows experimentation and diversity of platforms, with the best able to interconnect over time. The key societal question is whether people will tolerate a more federated social experience or gravitate back to the convenience of central platforms. The answer may depend on how seamless interoperability becomes.

\subsubsection{\textbf{Data Synchronization}}

Synchronizing data across decentralized platforms like \textit{Mastodon} is inherently unreliable. Actions such as likes, boosts, follows, and posts may not propagate consistently across servers, leading to discrepancies. A profile might show different follower counts on different instances~\cite{turn0search1}, or engagement on a post may appear low simply because updates haven’t arrived yet. These issues stem from the eventual consistency model most federated systems rely on. Servers must pull updates from each other, and delays or downtime can cause missed events or incomplete timelines. Users have reported visible lags—especially on large instances like \textit{mastodon.social}—where activity takes hours to appear elsewhere~\cite{turn0search1}. In P2P systems like \textit{SSB}, the situation is even more fragmented: each user sees only what their recent connections have shared. Until gossip protocols converge, views of the network remain subjective, leading to incomplete threads or misleading engagement signals. Such inconsistencies, while a trade-off of decentralization, can confuse users used to the synchronized experience of centralized platforms.

\subsection{\textbf{Operational Sustainability}}
Operational sustainability concerns the long-term viability of decentralized platforms, including how they incentivize maintenance, handle infrastructure costs, scale user activity, and resist degradation without funding.

\subsubsection{\textbf{Managerial Incentives}}
Sustaining decentralized platforms at scale requires aligning incentives with both economic viability and operational responsibility~\cite{Preston2025MastodonNonprofit, jeong2021fbadtracker, li2019incentivized}. For long-term success, managerial incentives should increasingly emphasize cost-efficiency and the creation of mechanisms that reward ongoing participation and governance. In federated systems, admins often act out of ideological or community-driven motivations, which, while fostering organic growth, can lead to burnout and financial strain without institutional support or monetization strategies. As platforms grow, these managerial roles must be incentivized through reliable funding. P2P networks introduce further challenges: without built-in incentives to store, relay, or maintain data, users are likely to behave in self-serving ways that degrade network performance. While reputation scores or token economies are common approaches, they remain fragile unless they balance incentives against cost-effectiveness and network contribution. Blockchain systems take a more explicit stance, offering rewards to validators and sometimes content contributors. However, the sustainability of such models depends on careful economic design and minimizing speculative distortions that can lead to inefficiencies or unintended behaviors.


\subsubsection{\textbf{Platform Growth Handling}}
As decentralized platforms scale, their long-term viability hinges on both user adoption and their ability to maintain acceptable performance levels. Platform growth strategies should focus on architectural scalability and minimizing response times to preserve user experience under load. Federated systems scale horizontally through the addition of new servers, but each node adds coordination and moderation complexity. Without centralized funding or operational support, servers often depend on volunteer labor, which can bottleneck scalability and reduce resilience~\cite{hoover2023mastodon}. Efficient onboarding, interoperability standards, and funding pathways are essential to supporting growth without overburdening individual instances. P2P architectures distribute infrastructure costs to end users, enabling serverless growth but often at the cost of reliability. Because availability and performance depend on active peer participation, such systems can exhibit high latency or churn unless a critical mass of users is consistently online. This model demands optimization for response time, lightweight protocols, and embedded incentives to sustain performance at scale. Blockchain-based social networks must navigate consensus-related constraints. Mechanisms such as proof-of-work and proof-of-stake introduce latency and significant computational costs, limiting throughput and increasing response times. To scale effectively, these systems often rely on off-chain solutions or hybrid architectures, which must be carefully designed to preserve decentralization while improving responsiveness. Overall, platform growth must prioritize technical scalability alongside architectural strategies that maintain performance as user bases expand.

\subsection{\textbf{Account Credibility}}
Account credibility concerns the challenge of verifying identity, assessing trustworthiness, and maintaining behavioral integrity, where central gatekeeping mechanisms such as verification are absent or constrained


\subsubsection{\textbf{Bot Accounts}}
Bot accounts in decentralized social networks can take many forms, from benign to malicious. On \textit{Mastodon}, account creation is often unrestricted on open servers, and users can even run their own servers dedicated entirely to bots (e.g., \texttt{botsin.space}, \texttt{feedsin.space}, etc). This flexibility has enabled the growth of creative and useful bots --- such as those posting art, quotes, or news --- but also contributed to excessive content generation and moderation burdens for unidentified bots~\cite{bono2024exploration}. What makes matters worse, it is easier for malicious bot operators to evade bans by just hopping to new servers or new accounts. Indeed, the low barrier to account creation means a persistent troll can create dozens of accounts across the federated networks with relative ease. Network-wide, distinguishing bot accounts is a challenge when data is spread out. Developers are focusing on bot-detection tools tailored to decentralized networks, but it is in the early stage~\cite{mastodon_v428_release}. Especially in P2P networks like \textit{Nostr}, the ease of generating new key pairs at no cost has led to widespread bot activity. To mitigate this, some relays have begun requiring proof-of-work or small Bitcoin fees as a barrier to posting.


\subsubsection{\textbf{Impersonation}}

The decentralized structure makes it easy for malicious actors to create accounts impersonating individuals or organizations, undermining trust. On \textit{Mastodon}, there is no universal username namespace. While that is similar to email, many users are not used to verifying which server account is on and may fall for lookalike handles on a different domain. For example, one could register \texttt{elonmusk@FakeServer} and pretend to be Elon Musk. \textit{Mastodon} and others have introduced profile verification by linking their personal, but many users do not check those. Recently, \textit{Bluesky} encountered impersonation challenges when a surge of fake celebrity accounts began appearing online~\cite{bluesky2024impersonator}. In response, the platform implemented policies aimed at addressing impersonation, primarily relying on user self-reporting and verification mechanisms~\cite{bluesky_moderation}. One such method allows users to verify their identity by linking their handle to a domain name they control. For example, the username \texttt{@JordanPeterson} will be considered verified if the user owns his official website \texttt{jordanpeterson.com}. There are also third-party services like \textit{ClearSky}\footnote{\url{https://clearsky.app/}} contribute by ranking users based on the number of blocks they receive, promoting community-driven identification of bad actors. However, the overall effectiveness of these decentralized approaches remains uncertain, especially when compared to moderation systems on centralized platforms. In blockchain-based networks like Lens or \textit{DeSo}, identity is linked to wallet addresses but remains pseudonymous. If a well-known individual wants to join, they might have to announce their wallet address through their blockchain address out-of-band to prove which account is actually theirs




\subsubsection{\textbf{Spamming}}

In federated networks, such as \textit{Mastodon}, spam typically originates from rogue servers that flood the network with unsolicited content --- ranging from mass follow requests to low-quality promotional posts. These actors hijack an abandoned server to breed spam into otherwise trusted networks~\cite{silberling2024discord}. While admins can respond by defeating malicious domains, this reactive defense is limited: spammers can rapidly deploy new instances, or poorly maintained servers to resume their activity~\cite{perez2024mastodon_spam}. P2P networks face even steeper anti-spam challenges. Without centralized infrastructure, each peer is responsible for moderating incoming content. Spammers can target a wide swath of peers directly, bypassing any notion of global filtering. Some systems employ trust graphs or web-of-trust models to prioritize content from known or reputable sources. However, these approaches can reduce network openness, hindering serendipitous or diverse interactions --- a hallmark of decentralized design. Blockchain-based platforms also show vulnerability to spam, especially when transaction costs are minimal. Low posting costs allow users to flood the network with repetitive, manipulative, or reward-seeking content. Platforms like \textit{Steemit} have seen coordinated spam campaigns that exploit token-based incentive systems. \textit{Steemit} have introduced microtransaction fees or staking requirements to deter such behavior, but these economic deterrents are often insufficient to eliminate spammers completely\cite{li2019incentivized}.

\subsection{\textbf{Behavioral Dynamics}}
Behavioral dynamics refer to the emergent patterns of user interaction, influence, and community formation in decentralized platforms, often mirroring or diverging from those observed in centralized social systems

\subsubsection{\textbf{Interaction Monopoly}}

Decentralization can inadvertently give rise to user behaviors that lead to re-centralization, where a small number of entities or individuals accumulate disproportionate influence—ultimately replicating the monopolistic dynamics these systems were designed to avoid. In federated systems such as \textit{Mastodon}, certain instances have emerged as dominant hubs. For example, \texttt{mastodon.social} became a focal point of user activity after Elon Musk’s acquisition of \textit{Twitter} in October 2022. In the month following the acquisition, Mastodon saw a surge in adoption, reaching approximately 850,000 active users on \texttt{mastodon.social}, which is the server managed by the founder of \textit{Mastodon}. This overload required an upgrade on the infrastructure on the server to resume its registration~\cite{raman2019challenges, la2021understanding}. This kind of concentration centralizes moderation authority and user interactions, potentially undermining the decentralized ethos of the network. Similar dynamics can be observed in P2P networks. While these systems remove centralized servers entirely, users often gravitate toward high-visibility nodes or influential figures for content and interaction. This behavior can create emergent hubs of influence, replicating hierarchical patterns found in centralized platforms. Blockchain-based social networks exhibit analogous trends. On platforms like \textit{Steemit}, users with substantial token holdings --- commonly referred to as “whales” --- can exert outsized influence over content visibility and reward distribution~\cite{sai2021taxonomy, de2019fragility}. Since voting power is tied to token ownership, these individuals can dominate curation mechanisms, discouraging broader participation and entrenching content monopolies.

\subsubsection{\textbf{Echo Chamber Effect}}

Decentralized platforms such as \textit{Mastodon}, \textit{Bluesky}, and others are increasingly vulnerable to echo chambers and ideological polarization due to their fragmented and user-governed architectures~\cite{la2024polarization, di2024characterizing}. Without centralized moderation, communities often self-select into ideologically homogenous groups, leading to reduced exposure to dissenting views~\cite{oxford2024social}. This effect can intensify especially when users migrate following deplatforming events. These migrations can create friction between existing communities and incoming groups, especially when ideological values clash~\cite{monti2023online, thomas2023disrupting}. Over time, such tensions may lead to stronger, more insular communities that are harder to engage in open public discourse, particularly when platform structure limits message visibility across the broader network. For example, in 2024, a reporter, Jesse Singal, was aggressively targeted on \textit{Bluesky} for his views on anti-trans views, prompting waves of harassment and blocking by users who opposed his stance~\cite{, Perez2024Bluesky}. The event demonstrated how decentralized spaces can be a hub for group-based antagonism, making it difficult to challenge dominant narratives within tightly bound clusters~\cite{corradini2024deconstructing}.


\subsubsection{\textbf{External Influence}}

Decentralized platforms can inadvertently influence mainstream social media by providing refuge for content that is moderated or censored elsewhere. Communities seeking to circumvent content restrictions may migrate to these networks, using them to coordinate or disseminate material that eventually circulates back into mainstream platforms. A notable example is the alt-tech social platform \textit{Gab}, which forked \textit{Mastodon} in 2019 to create its own minimally moderated instance~\cite{Guardian2024AltPlatforms}. Social media researchers highlight that adaptive links can spread through online hate networks --- such as the strategic sharing of \textit{Gab}'s content with links to mainstream platforms like \textit{Twitter} and \textit{YouTube} --- allow these communities to extend their reach and influence beyond the boundaries of the decentralized platforms themselves~\cite{zheng2024adaptive}. Moreover, centralized platforms are increasingly engaging with decentralized platforms. For example, \textit{Threads}~\cite{su2024threads} enabled cross-platform interactions with decentralized networks like \textit{Mastodon} through its “Fediverse Sharing” feature~\cite{jeong2025fediversesharing}. Addressing the complex challenges posed by this interaction of centralized and decentralized platforms requires coordinated action to monitor and mitigate the spread of harmful content on the social media ecosystem.

\section{\textbf{Conclusion and Future Work}}
Decentralized Online Social Networks (DOSNs) have evolved rapidly over the past decade, transcending experimental frameworks to become prominent platforms with substantial user bases, driven in part by recent migrations from centralized services. This study provides a comprehensive exploration of DOSN architectures including federated, peer-to-peer, blockchain-based, and hybrid models. We highlight their developmental trajectories, technical trade-offs, and societal implications when these platforms are deployed to the public.

Our paper reveals that while decentralized architectures offer notable benefits --- such as enhanced user autonomy, data sovereignty, and resistance to censorship --- they also introduce persistent technical and societal challenges. Scalability, data persistence, privacy protections, and administrative limitations continue to pose substantial hurdles, varying distinctly across different architectural models. Furthermore, these technical choices inevitably influence societal outcomes, including platform safety, user credibility, and community dynamics.

Looking ahead, the effective advancement of DOSNs requires careful balancing between decentralization's inherent advantages and its accompanying complexities. Future research and platform development should focus on key challenges. These include building moderation tools that work without central control~\cite{agarwal2024decentralised, jeong2022nothing, jeong2022classifying}, improving how systems scale and perform over time, enabling communication between different networks, and supporting long-term sustainability through effective incentive design. Addressing these areas is essential to shaping decentralized platforms that are technically viable, socially responsible, and sustainable while aligning with user expectations and societal norms.

.

\begin{acks}
This research was supported by the following grant: MURI: Persuasion, Identity, \& Morality in Social-Cyber Environments; Office of Naval Research Grant No. N000142112749. The views and conclusions are those of the authors and should not be interpreted as representing the official policies, either expressed or implied.
\end{acks}


\bibliographystyle{ACM-Reference-Format}
\bibliography{references}


\end{document}